\documentclass[12pt]{iopart}
\newtheorem{lemma}{Lemma}

\newtheorem{theorem}{Theorem}

\usepackage{iopams}
\usepackage{graphicx,amssymb}
\usepackage{color}

\usepackage{ifthen}

\begin{document}

\title[On Newtonian frames]
{On Newtonian frames}

\author{Bartolom\'{e} Coll$^1$, Joan Josep Ferrando$^2$\
and Juan Antonio Morales$^2$}

\address{$^1$\ Syst\`emes de r\'ef\'erence relativistes, SYRTE-CNRS,
Observatoire de Paris, \\75014 Paris, France.}

\address{$^2$\ Departament d'Astronomia
i Astrof\'{\i}sica, Universitat de Val\`encia, \\E-46100 Burjassot,
Val\`encia, Spain.}

\ead{bartolome.coll@obspm.fr; joan.ferrando@uv.es;
antonio.morales@uv.es}

\begin{abstract}
In Newtonian space-time there exist four, and only four, causal
classes of frames.
Natural frames allow to extend this result to coordinate systems, so
that coordinate systems may be also locally classified in four
causal classes.
These causal classes admit simple geometric descriptions and
physical interpretations.
For example, one can generate representatives of the four causal
classes by means of the  {\em linear synchronization group}.
Of particular interest is the {\em local Solar time
synchronization}, which reveals the limits of the frequent use of
the  concept of `causally oriented coordinate', such as that of
`time-like coordinate'.
Classical {\em positioning systems}, based in sound or light
signals, are, by themselves, interesting examples of location
systems, i.e. of physically constructible coordinate systems.
They show that one can locate events in Newtonian space-time {\em
without} any use of the concept of synchronization. In fact, the
coordinate systems associated to positioning systems, belong to all
the classes but the standard one, i.e. the one based in the
simultaneity synchronization.
The relativistic analogs of these examples, emphasize the contrast
between the four Newtonian and the one hundred and ninety nine
Lorentzian causal classes of frames of classical and relativistic
space-times, respectively.
\end{abstract}

\pacs{0420-q, 45.20.Dd, 0420Cv, 9510Jk}



\section{Introduction}
\label{intro}

Location systems are physical realizations of coordinate systems.
From laboratory do\-mains, Earth surface physics or global
navigation systems to space physics, solar system or celestial
astronomy, location systems allow the explicit construction of the
correspondence between the events of the observable physical world
and the points of  its mathematical space-time model in the physical
theory in use.

A location system must include the protocols for the physical
construction of the coordinate lines, coordinate surfaces or
coordinate hypersurfaces of the coordinate system that it physically
realizes. Thus, for example, these coordinate elements may be
realized, among other ways, by means of clocks for timelike lines,
laser pulses for null lines, synchronized inextensible threads for
spacelike lines, inextensible threads or laser beams for time like
surfaces, light-front signals for null hypersurfaces and so on. The
point of interest here is that every protocol physically realizes
coordinate lines, coordinate surfaces or coordinate hypersurfaces of
specific {\em causal orientations}. Conversely, the causal
orientation of the ingredients of a coordinate system intimately
constraints the physical protocols needed for the construction of
the corresponding location system.

The different protocols involved in the construction of location
systems give rise to coordinate elements (lines, surfaces and
hypersurfaces) of different causal orientations, i.e. they realize
coordinate systems of different causal nature. It is known that the
number of coordinate systems of different causal nature that can be
constructed in relativistic space-times is of exactly one hundred
and ninety nine \cite{199}. But the corresponding question for the
Newtonian space-time has never been asked until recently
\cite{Escola}.

Here this question is analyzed and it is shown that, in strong
contrast with the relativistic case, the number of Newtonian
coordinate systems of different causal nature reduces drastically to four.

A precise geometric description of these four classes is given and
some possible physical realizations of every one of them are
commented. Also, some examples are constructed of coordinate
systems for every one of these causal classes. And finally the four
causal classes of Newtonian coordinate systems are contrasted with
the one hundred and ninety nine Lorentzian causal classes and, among
them, specifically with their four relativistic analogs.

\subsection{Interest and applications of the causal classification of frames}
\label{subsecIntroInteresAndAppl}

The interest of the causal classification of coordinate systems is
not only taxonomic.

So, for example, in a similar way as three-dimensional Cartesian
coordinates frequently induce or are induced by a floor plan and
elevation cut of the space, every four-dimensional coordinate system
may be seen as a specific cut or foliation of (a region of) the
space-time in particular pieces: those defined by the coordinate
hypersurfaces, surfaces or lines of the coordinate system. But now
these cuts or foliations may be of different specific causal
classes. In this sense, the well known usual coordinate systems,
essentially based in a three-space foliation plus a one-time
congruence, are induced by, or induce, the standard {\em evolution
conception} of Newtonian and relativistic physics. But other cuts or
foliations, among the other three possible cuts or foliations in
Newtonian theory or among the other one hundred and ninety eight
possible cuts or foliations in relativity, may help us to better
describe and understand other aspects of the space-time, and even to
wake up our interest for variations of physical fields other than
the timelike ones, intimately induced by the evolution conception.

But perhaps the most imminent interest of the causal classification
of coordinate systems is appearing in the at present methods for
solving practical relativistic problems. Relativity theory is
conceptually considered  as a physically autonomous theory, i.e. a
theory that, for its development, needs no other physical concepts
that the ones contained in its specific foundations, or those that can be
coherently deduced from them. But in practice, in spite of the
efforts made in this direction
\cite{Escola, coll, bahder, Rovelli, blago, rey, 2D-A, 2D-B, 4D}, the
development of the least physical practical application needs, for
the moment, a detour to Newtonian concepts and post-Newtonian
methods. This situation reduces
relativity theory, up to little exceptions, to the role of a
{\em corrective algorithm} for Newtonian theory, relegating its best
specific concepts to a simple historically astute, but
otherwise ineffective, method of setting the main equations of the
theory, the Einstein equations. In fact, irrespective of the
revolutionary and paradigmatic concepts that general relativity
opposed to the Newtonian scope of the space-time, only quantitative
first terms in Taylor
development of Einstein equations with respect to a Newtonian
background remain essentially the unique element of general
relativity used to improve Newtonian results obtained under
Newtonian concepts.

As long as this situation remains, it is highly convenient in
post-Newtonian developments to choose location or coordinate systems
such that their causal properties be the same both for the
relativistically corrected metric structure as well as for the
starting Newtonian one. Otherwise, in going from Newtonian to
relativistic results by the addition of higher corrective terms,
one would add, to the quantitative corrective process involving
the physical quantities of the problem, qualitative corrections due
to an eventual change of  causal orientations of the coordinate elements
of the location system. If such a change takes place, the
physical interpretation of the vector or tensor {\em components}
of the physical quantities of the
problem, and therefore the adequate instruments for their measure,
could change drastically%
\footnote{Think that, for example, of
the four-dimensional energy tensor, the usual interpretation of
their components in terms of energy density, momentum density
and stress quantities is only valid for {\em standard} frames.
Standard frames  privilege {\em one} observer among all others,
but constitute a little class among the one hundred and ninety
nine classes of possible frames; in all the others, and in
particular in the real null frames of emission coordinates
(see below in the text), such an interpretation fails, because
no observers are necessary at all.}. %

Fortunately this convenient choice of analogous causal classes has
been made up to now, naturally but unconsciously. Simply because the
starting Newtonian coordinate system has been essentially chosen to be
the Cartesian one, and that the weak gravitational fields usually
considered in astronomy have been unable to change, with the lower order
perturbed relativistic values of the metric, their causal orientation.
But new problems, concerning black holes, binary systems, gravitational
waves, positioning systems, formation flight satellites and space physics,
could induce to start from other Newtonian coordinate systems, best
adapted to these problems or to push away higher order terms. And
then, changes in the causal orientation of some of the ingredients
of the starting Newtonian coordinate system become possible when
evaluated with the corrective algorithm generating the relativistic
space-time metric.

In fact, in numerical relativity, a verification not only of the
regularity but of the stability (constancy) of the whole causal
class of the coordinate system would be also convenient in order to
guarantee the physical interpretation, at least, of the components
of the energetic quantities present in Einstein equations.

These are the main points of interest involving related causal
classes of Newtonian {\em and} relativistic coordinate systems.
Other points of interest concerning specifically relativistic
coordinate systems were mentioned in \cite{199}.

But, in order to better understand the role that location systems as
physical objects, or coordinate systems as mathematical objects,
play in the conception and analysis of experimental situations, a
lot of work remains to be done, the present one being only one of
the first little pieces. Recently considered {\em emission
coordinates} go in this direction (see \cite{2D-A, 2D-B, 4D} and
references therein).

\subsection{Structure of the present work}
\label{subsecIntroPresentWork}

The paper is organized as follows.  In Sec. \ref{sec:2} the notion
of causal class of a frame is introduced and extended to coordinate
systems. Sec \ref{sec:3} characterizes the four causal classes of
frames or coordinate systems in Newtonian space-time, and extends
this result to arbitrary dimension. In Sec. \ref{sec:4} the notions
of coordinate parameter and gradient coordinate are emphasized in
order to better understand the limits of the assignation of a causal
character to the coordinates, and the first elements of the
synchronization group are stressed for the incoming applications.
Sec. \ref{sec:5} presents some physical examples of Newtonian
coordinates of the four causal classes. It is shown that the linear
synchronization group is able to generate coordinate systems of {\em
any} of the four causal classes, the causal class of the ancestral
local Solar time is obtained and commented, and Newtonian emission
coordinates generated by positioning systems, able to locating
events out of any notion of synchronization, are shown to belong to
any causal class but the usual one. In Sec. \ref{sec:6} Newtonian
and Lorentzian classes are contrasted across the relativistic
analogs of the chosen Newtonian examples. Finally, in Sec.
\ref{sec:7} we comment on the role that our results can play as
training toys for a better understanding of the physical space-time.

Some preliminary results about this work were presented as a
contributing lecture at the school on {\it Relativistic Coordinates,
Reference and Positioning Systems} \cite{Escola}.

\section{Notion of causal class}\label{sec:2}

In relativity, directions and planes or hyperplanes of directions at
an event are said to be spacelike, null or timelike {\em oriented}
if they are respectively exterior, tangent or secant to the
light-cone of this event. These {\em causal orientations}, of clear
geometrical and physical meaning, extend naturally to vectors and
volume forms on these sets of directions.

Thus, every one of the vectors $v_A$ of a frame $\{v_A\}$\, $(A =
1,...,4)$ has a particular causal orientation ${\rm c}_A\,.$ What
about the causal orientations ${\rm C}_{AB}$\, $(A<B)$ of the six
{\em associated} planes $\Pi(v_A,v_B)$ of the frame? Are they
determined by the sole causal orientations ${\rm c}_A$ of the
vectors of the frame? Certainly not, because for example the plane
associated to two spacelike vectors may have any causal
orientation. So, in general, the specifications ${\rm c}_A$ and
${\rm C}_{AB}$ are independent.

Moreover, in order to give a complete description of the causal
properties of the frames, one needs also to specify the
causal orientations $\it{c}_A$ of the four covectors $\theta^A$
giving the dual frame $\{\theta^A\}$, $\theta^A (v_B) = \delta^A _B$.
The $\it{c}_A$'s are one-to-one related to the causal orientations of
the four associated 3-planes $\Pi(v_B, v_C, v_D)$ with
$\theta^A(v_B) =$ $ \theta^A(v_C) =$ $ \theta^A(v_D) = 0$ which are
not determined, in general, by the specification of both ${\rm c}_A$
and ${\rm C}_{AB}$.

The set of $(4+6+4=)$ $14$ causal orientations $\{\rm c_A, \rm
C_{AB}, \it{c}_A\}$ is called the {\em causal signature} of a frame
$\{v_A\}$, and characterizes completely its {\em causal class}: the
causal class of a frame is the set of all the frames that have
same causal signature. The causal signature of a frame provides
exhaustive information about the causal properties of its geometric
elements (directions, planes and hyperplanes). Elsewhere \cite{199},
the following result was obtained.
\begin{theorem} \label{teo-199}
In a four-dimensional Lorentzian space-time there exist 199 causal
classes of  frames.
\end{theorem}

As a {\em natural frame} is nothing but the set of derivations along
the parameterized lines of a coordinate system, the notion of causal
class extends naturally to the set of coordinate lines of the
coordinate system and so, to the coordinate system itself. But
because this extension of the notion of causal class to a coordinate
system is by construction a point by point extension, i.e. the
causal class of a coordinate system is the causal class of its
natural frame at every point, a coordinate system may present
different causal classes at different points of its domain of
definition. Indeed, some examples of this situation will be given
below.

The assignment of one specific causal class to a coordinate system
in a region of the space-time supposes that the causal orientations
of all the geometric elements of the coordinate system (lines,
surfaces and hypersurfaces) are the same at any point of the region
or, in other words, that the region under consideration is a {\em
causal homogeneous region} for the coordinate system in question.

Theorem \ref{teo-199} equivalently states that there are 199
causally different ways to parameterize the events of a relativistic
space-time causal homogeneous region. The complete and explicit
specification of them was given in \cite{199} and  more recently in
\cite{Escola}.

By definition, the causal class of a coordinate system
$\{x^\alpha\}_{\alpha=1}^4$ in a domain is the causal class $\{\rm
c_\alpha, \rm C_{\alpha\beta}, \it{c}_\alpha\}$ of its associated
natural frame at the events of the domain. The ${{\rm c}_\alpha}$'s
are the causal orientations of the vectors $\partial_\alpha \equiv
\displaystyle{\frac{\partial}{\partial x^\alpha}}$ of the natural
frame $\{\partial_\alpha\}$ itself, and the  ${\it c_\alpha}$'s are
the causal orientations of the $1$-forms $d x^\alpha$ of the coframe
$\{d x^\alpha\}.$ Four families of coordinate $3$-surfaces
(hypersurfaces) are associated with this coframe, and their mutual
intersections give six families of coordinate $2$-surfaces
(surfaces) whose causal orientations are precisely given by ${\rm
C_{\alpha \beta}}$ (of course, the mutual intersections of these
surfaces give the four congruences of coordinate lines of causal
orientation ${\rm c_\alpha}$). We have chosen the following order
for the causal orientations of a causal class:  $\{{\rm c_1 c_2 c_3
c_4}, {\rm C_{12}C_{13} C_{14} C_{23} C_{24} C_{34}}, {\it c_1 c_2
c_3 c_4}\}.$

What is the situation in Newtonian physics concerning causal
orientations and causal classes? Of course, now the causal
orientations ${\rm c_A},$ ${\rm C_{AB}},$ ${\it c_A}$ reduce to be
only of timelike or spacelike character. But a causal class needs
also to be characterized by the fourteen quantities $\{{\rm c_A},$
${\rm C_{AB}},$ ${\it c_A}\}.$ Nevertheless now some of them
determine systematically the others. Specifically, we shall show in
Section 3 that for Newtonian frames one has the implications
$$
    \{{\rm c_A}\}\Rightarrow \{{\rm C_{AB}}, {\it c_A}\}  \quad ,
    \quad  \{{\rm C_{AB}}\}\Rightarrow \{{\it c_A}\} \quad ,
$$
but
$$
    \{{\rm C_{AB}}\}\nRightarrow \{{\rm c_A}\} \quad ,
    \quad \{{\it c_A}\}\nRightarrow \{{\rm c_A}, {\rm C_{AB}}\} \quad .
$$

These implications lead to a Newtonian situation remarkably simpler
than the Lorentzian one. In fact, surprisingly enough at first
glance, only four causally different classes of frames or coordinate
systems are admissible in Newtonian space-time (see Sec. \ref{sec:3}
below). It is startling that, in spite of this poverty of classes,
only the  {\em standard class} (i. e. the one wholly adapted to the
absolute space $\oplus$ time Newtonian decomposition) has been
explicitly referred to in the literature. In the next section we
construct these four classes of Newtonian frames.

%
%

\section{Causal classes of Newtonian frames}
\label{sec:3}

The differences in the geometric description of Lorentzian and
Newtonian frames come from the causal structure induced by the
metric description of the underlying physics.

In Relativity the space-time metric defines a one-to-one
correspondence between vectors and covectors at every event. In
contrast, in Newtonian physics no non-degenerate metric structure
exists. The degenerate metric structure is given by a rank one
covariant positive {\em time metric} $T$ and an orthogonal rank
three contravariant positive {\em space metric} $\gamma^*,$ $T
\times \gamma^*   = 0,$ where $\times$ stands for the cross
product%
\footnote{The cross product $\times$, or matrix product, is the
contraction of the adjacent vector spaces of the tensor product
$\otimes$. In tensor components, $T \times \gamma^* $ is written as
$T_{\alpha\rho}\gamma^{*\rho\beta}$.}.

The time metric $T$ is necessarily of the form $T = \theta \otimes
\theta,$ where the 1-form $\theta,$ the {\em time current,} defines
the unit of time. That this time is {\em uniform} for any observer,
or {\em absolute}%
\footnote{Absolute and uniform times are strongly
related. See \cite{UT}.},
implies the exact character of the time
current, $\theta = dt,$ where $t$ is any absolute time
scale\footnote{A time scale is a rhythm generated by a unit interval
together with a choice of origin.}. The hypersurfaces $t =$ {\em
constant} constitute the {\em instantaneous spaces},
{\em simultaneity loci} or {\em spaces} at the instant $t.$

It should be stressed that the above elements, $T$ (or
$\theta$) and $\gamma^*$, already determine the Newtonian causal
structure%
\footnote{Nevertheless, for the formulation of the equations of
motion, a flat and symmetric affine connection is also required in
order to introduce inertia. In addition, in the four-dimensional
formulation of Newtonian gravity, the requirement of another
symmetric, non-flat and not metric connection is needed in order to
introduce the gravitational field \cite{Cartan, UT, Kil,
Trautman1,Trautman2,Trautman3}, but we shall not need  them in this
work.}
Here, we are interested only in the causal orientation at every
event of directions, planes and hyperplanes induced by the sole
Newtonian structure provided by $\theta$ and $\gamma^*$. In this
structure, a vector $v$ is {\em spacelike} if it is instantaneous
with respect to the time current $\theta$, i.e. if $\theta(v) = 0$.
Otherwise, the vector is {\em timelike}. A timelike vector $v$ is
{\em future} (resp. {\em past}) {\em oriented} if $\theta(v)>0$
(resp. $\theta(v)<0$). Obviously, these notions apply naturally to
vector fields in causal homogeneous regions.

It is clear that a basis can have {\it at most} three spacelike
vectors so that, denoting with Roman letters ($\rm{e, t}$) the
causal orientations (respectively spacelike, timelike) of vectors,
it holds:
\begin{lemma}\label{lem1} Attending to the causal
orientation of their vectors, there exist four causal {\em types} of
Newtonian bases, namely: $\{\rm{t e e e} \},$ $ \{\rm{t t e e} \},$
$ \{\rm{t t t e} \},$ $ \{\rm{t t t t} \}.$
\end{lemma}

In a Newtonian structure, correspondingly, a covector $\omega \neq
0$ is {\em timelike} if it has no instantaneous part with respect to
the space metric $\gamma^*$, i.e. if $\gamma^*(\omega)=0$.
Otherwise, the covector $\omega$ is {\em spacelike}.  The sole
timelike codirection is that defined by the current $ \theta$ at
every event because $\gamma^*$ has rank $3$. Thus, if $\omega$ is
timelike it is necessarily of the form $\omega = a \, \theta$ with
$a \neq 0$. Then $\omega$ is {\em future} (resp. {\em past}) {\em
oriented} if $a>0$ (resp. $a<0$). Obviously, these notions are also
naturally valid for 1-forms in causal homogeneous regions.

It is then clear that a cobasis has {\it at most} one timelike
covector so that, denoting with Italic letters ($\it{e, t}$) the
causal orientations (respectively spacelike, timelike) of covectors,
it holds:
\begin{lemma}\label{lem2}
Attending to the causal orientation of their covectors, there exist
two causal {\em types} of Newtonian cobases, namely: $\{{\it t e e
e}\}, \{{\it e e e e}\}$.
\end{lemma}

Lemmas \ref{lem1} and \ref{lem2} show the lack of symmetry of causal
types of Newtonian bases and cobases, in contrast to the rigorous
symmetry of the relativistic case.

A $r$-plane $\Pi$ is {\em spacelike} if every vector $v$ in it is
spacelike. Otherwise, $\Pi$ is {\em timelike}, i.e. it
contains timelike vectors. Two (resp. three) linearly independent
spacelike vectors generate a spacelike $2$-plane (resp.
$3$-plane).

A $r$-coplane $\Omega$ is {\em timelike} if it contains the time current
$\theta$. Otherwise $\Omega$ is {\em spacelike}.

The annihilator coplane ${\Omega}_\Pi$ of a $r$-plane $\Pi$ is the
$(4-r)$-coplane
$${\Omega}_\Pi \equiv \{\omega \, | \,  \omega(v)= 0 \,\,\,\, \forall
v\in \Pi\}.$$ Obviously, these definitions apply also to $r$-plane fields and
$r$-coplane fields in causal homogeneous regions.

Accordingly, we have the following result.
\begin{lemma}\label{lem3}
A r-plane $\Pi$ is spacelike {\em (resp.} timelike{\em )} iff
$\Omega_\Pi$ is timelike {\em (resp.} spacelike{\em ).}
\end{lemma}

In particular, given a Newtonian frame $\{v_1, v_2, v_3, v_4\}$, a
covector $\theta^\alpha$ of its {\it dual} frame
$\{\theta^1,\theta^2, \theta^3, \theta^4\}$ is timelike
(resp. spacelike) iff the $3$-plane generated by $\{v_\beta \}_{\beta
\neq \alpha}$ is spacelike (resp. timelike).

On account of the above considerations, the causal orientations of
the four vectors of a Newtonian frame determine unambiguously the
causal orientations of their six associated 2-planes and the causal
orientations of their four associated 3-planes. Consequently, we reach
the following result.
\begin{theorem}\label{teoNw}
In the $4$-dimensional Newtonian space-time there exist four, and
only four, causal classes of frames.
\end{theorem}
\begin{figure}
\begin{flushright}
\parbox[c]{0.84\textwidth}{\includegraphics[width=0.84\textwidth]{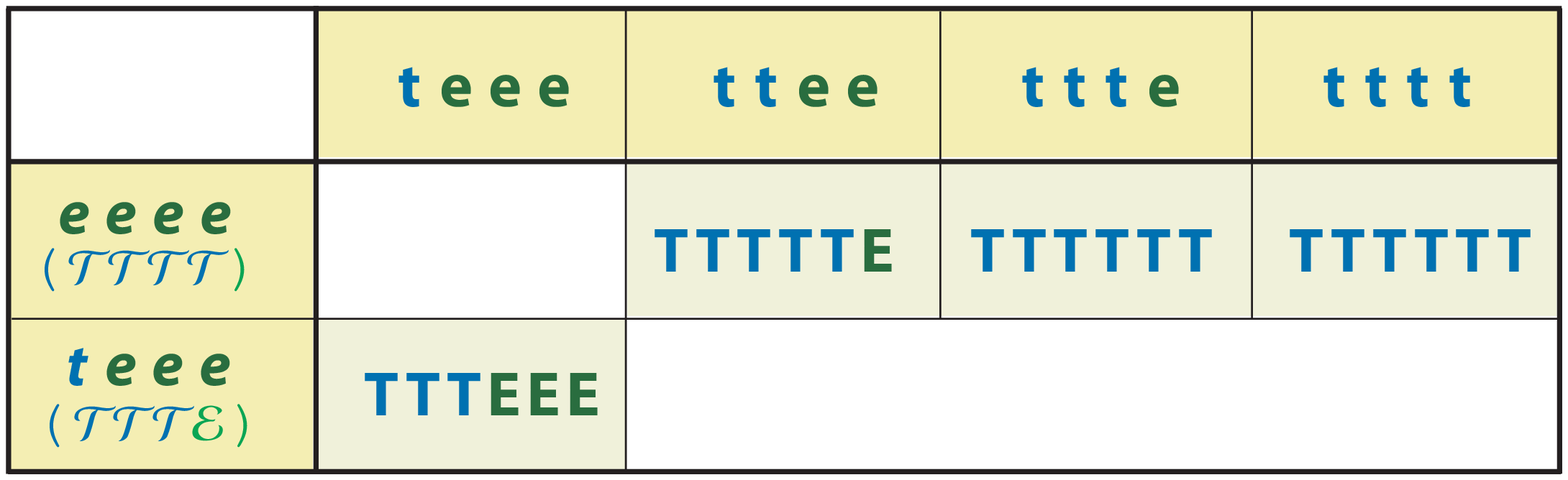}}
\end{flushright}
\caption{The four causal classes of Newtonian frames. Roman letters
($\rm{e},\rm{t}$), capital letters ($\rm{E, T}$), calligraphic
($\cal{E, T}$) and Italic ($\it{e, t}$) letters represent the causal
orientations (spacelike, timelike) respectively of the vectors of
the frame, of their associated 2-planes, of their associated 3-planes
and of the covectors of the coframe. This causal classification
extends naturally to coordinate systems in causal homogeneous
regions.
\label{New-clases}}
\end{figure}

The four Newtonian causal classes are represented in Fig.
\ref{New-clases} whose reading is as follows.
\begin{enumerate}
\item The first column shows the sets of causal orientations
$c_A = \{e\, e\, e\, e\},$ $c_A = \{t\, e\, e\, e\}$ of the covectors
of the coframe (or correspondingly, of the sets of causal orientations
$\bar{c}_A = \{{\cal{T T T T}}\},$ $\bar{c}_A = \{{\cal{T T T E}}\}$
of the four 3-planes of the frame or of the four families of coordinate
hypersurfaces of a coordinate system). As stated in Lemma \ref{lem2},
only these two sets are possible, up to permutations.
\item The first file shows the sets of causal orientations
$\rm{c}_A = \{{\textrm t\, e\, e\, e }\}$, $\rm{c}_A = \{{\textrm t\,
t\, e\, e }\}$, $\rm{c}_A = \{{\textrm t\, t\, t\, e }\}$, $\rm{c}_A =
\{{\textrm t\, t\, t\, t }\}$ of the vectors of the frames or,
correspondingly, the sets of causal orientations of the congruences of
coordinate lines of a coordinate system. As stated in Lemma \ref{lem1},
only four sets are possible, up to permutations.
\item Each not empty $(p, q)$-cell $(p\!=\!1, 2;\ q\!=\!1, 2, 3, 4)$
shows the set of causal orientations ${\rm C}_{AB}$ of the associated
 $2$-planes of vectors of the $q$-th frame, that corresponds to
the $p$-th coframe or, correspondingly, the set of causal orientations of
the six coordinate surfaces of a coordinate system.
\item Permutations of the vectors of the frame or of the covectors
of the coframe  induce permutations of the associated 2-planes and
3-planes, but do not alter their causal class. Correspondingly,
permutations of the lines or hypersurfaces of a coordinate system
induce permutations of the coordinate surfaces of the system, but do
not alter its causal class.
\end{enumerate}

For instance, standard frames, i.e. those that are locally realized
with three rods and one clock at rest with respect to the rods,
belong to the causal class $\{{\rm t e e e, T T T E E E}, {\it
teee}\}.$  The history of the clock is a timelike coordinate line.
The other coordinate lines are spacelike straight lines tangent to
the rods at every (clock's) instant.

Geometrically, this causal class is better visualized  by the family
of spacelike instantaneous 3-planes generated by the directions
of the three rods  and the three families of timelike 3-planes
(each one being the history of the 2-plane generated by two rods),
whose normals or algebraic duals  define the natural coframe  $\{t e
e e\}$. The mutual cuts of these coordinate 3-planes give the six
families of coordinate 2-planes (denoted $\{\rm{T T T E E E}\},$ three
of them being timelike and the other three ones being spacelike).
The coordinate planes cut in four congruences of coordinate lines
(now denoted $\{\rm{t e e e}\},$ one being timelike and the others
being spacelike).

As already mentioned, the simplicity of the  Newtonian causal
structure with respect to the Lorentzian one lies in that the causal
type of a Newtonian frame determines completely its causal class.
This is related to the fact that, in Newtonian space-time, any set
of spacelike vectors always generates a spacelike subspace. As a
consequence, the number of causally different Newtonian classes of
frames is equal to the dimension of the space. This is a general
property, independent of the dimension $n$ of the space-time. Denoting
by $\{k\,{\rm t}, (n-k)\,{\rm e}\}$ the causal type of a basis with
$k$ timelike vectors and $n-k$ spacelike ones%
\footnote{The comma between different causal orientations is put in
this condensed expression only for visual clarity.},
we therefore have:
\begin{theorem}\label{teoNwn} In the $n$-dimensional Newtonian
space-time there exist $n$ causal classes of frames. A basis whose
causal type is $\{k\,{\rm t}, (n-k)\,{\rm e}\},$ $k=1, ..., n,$ has
 ${n-k \choose r}$ \, spacelike  associated $r$-planes and ${n
\choose r} - {n-k \choose r}$ timelike associated $r$-planes
$(r= 1, ..., n)$.
\end{theorem}
In dimension $n$, the causal classification of Newtonian frames in
$n$ classes induces a causal partition of the general lineal group
$GL(n).$ Like in the Lorentzian case, the {\em restriction} of
$GL(n)$ to a sole of these partitions simplifies notably the study
of {\em intrinsic} deformations or perturbations of metric
structures. In other, more intuitive, words, when one performs an
arbitrary deformation of a metric structure, one obtains a mixed
result: a wanted variation of the metric structure itself and a
superfluous variation of the fields of frames (gauge) with respect
to which the metric is expressed. Our causal classification allows
us to reduce the group of deformations by considering its ``quotient''
by the causal classes, that is to say, roughly speaking, by
considering nothing but the n-th part of the group which
transforms metric structures but respects the causal class of the
field of frames in which they are expressed. But this aspect will be
analysed elsewhere.

In what follows, we will construct some examples of transformations
of $GL(n)$ that change the causal class of a starting coordinate
system and also we will give direct examples of coordinate systems
of the unusual causal classes. But previously we need to specify
some simple but important notions.

\section{Coordinate parameters, gradient coordinates and synchronizations}
\label{sec:4}

Whatever be the complete description of a coordinate system, it may
be equivalently determined by its coordinate hypersurfaces, that is
to say, by the four one-parameter families of hypersurfaces whose
mutual cuts give the six families of coordinates surfaces, which in
turn cut in the four congruences of coordinate lines.

Conversely, when the coordinate system is already know, say
$\{x^\alpha\}$, these geometric elements may be easily discerned:
the four one-parameter families of coordinate hypersurfaces are
given by $\{ x^\alpha = \mbox{constant} \}$, the six two-parameter
families of coordinate surfaces are given by $\{x^\alpha =
\mbox{constant}, x^\beta = \mbox{constant} \}$, and the four
three-parameter families of coordinate lines are given by
$\{x^\alpha = \mbox{constant}, x^\beta = \mbox{constant}, x^\gamma =
\mbox{constant}\}$ for superscripts $\alpha$, $\beta$, $\gamma$ such
that $\alpha$ $\neq$ $\beta$ $\neq$ $\gamma$ $\neq$ $\alpha$.

What Fig.  \ref{New-clases} shows is nothing but the four
possibilities of causal orientation of these geometric elements in
Newtonian space-time. Thus, for example, the class $\{{\rm t t t e,
T T T T T T}, {\it e e e e}\}$ represents those coordinate systems
whose four coordinate hypersurfaces are all timelike $\{\mathcal{T T
T T}\}$, cut in six families of timelike coordinate surfaces
$\{\rm{T T T T T T}\}$, which in turn cut in four congruences of
coordinate lines $\{{\rm t t t e}\}$, three of them timelike and the
other one spacelike.

\subsection{Coordinate parameters and gradient coordinates}
\label{subsec4CooParamGradCoo}

In fact, in any space-time, every coordinate $x^\alpha$ plays two
extreme roles: that of a (coordinate) hypersurface for every
constant value, of gradient $dx^\alpha$, and that of a (coordinate)
line when the other coordinates remain constant, of tangent vector
$\partial_\alpha$. This simple fact shows that, in spite of our
deep-seated custom of associating to a coordinate a causal
orientation, saying that {\em it is} timelike,  lightlike or
spacelike, {\em this appellation is not generically coherent.}
Causal orientations are generically associated with directions or
sets of directions of geometric objects, but not with space-time
variables or parameters associated to them. In the case of a
coordinate $x^\alpha$, this generic incoherence appears because its
two natural variations in the coordinate system, $dx^\alpha$ and
$\partial_\alpha$, have {\em generically different} causal
orientations. {\em Only} when both causal orientations coincide, it
is conceptually clear to extend to $x^\alpha$ itself the appellation
of the  common causal orientation of its two mentioned variations.

Consequently, we shall say generically of a coordinate $x^\alpha$
that it is a $c_\alpha$ {\em gradient coordinate} and a $\rm{c}_\alpha$
{\em coordinate parameter} when the causal orientations of its variations
$dx^\alpha$ and $\partial_\alpha$  be respectively $c_\alpha$ and
$\rm{c}_\alpha$.

In addition, of a coordinate $t$ which is a timelike coordinate
parameter and a timelike (resp. spacelike) gradient coordinate, we
shall say also that it defines a {\em spacelike} (resp. {\em
timelike}) {\em synchronization}  (the coordinate hypersurfaces $t =
\mbox{constant}$ being the synchronous event loci of the coordinate
lines $t = \mbox{variable}$. See below).

 It is to be noted that the appellation ``timelike coordinate parameter''
in place of the usual ``timelike coordinate'' when $t$ is also a
timelike synchronization is the correct one, because in that case
$t$ may be a constant or even a decreasing parameter along future
oriented timelike trajectories of the space-time coordinate region,
an odd property for a ``time coordinate''.

A paradigmatic example of this situation is the oldest timelike
coordinate parameter known by humanity, the {\em local Solar time},
that will be considered in Section \ref{sec:5}. But before analyzing
it, it is worthwhile to first present the group of (pure)
synchronizations and its finite dimensional subgroup, the group of
(pure) linear synchronizations.

\subsection{The Synchronization Group}
\label{subsec4SyncGr}

Consider a set of clocks in some region of a space-time. Their
histories constitute a set of timelike lines on the region,
naturally parameterized by the time $t$ of the clocks. A {\em
synchronization} is the stipulation of the locus of events where the
clocks display the time $t = t_0$ for some chosen constant value
$t_0$.

We are interested here for `smooth situations', in which the
smallness of the clocks, their number and their histories  are such
that they can be efficiently described by a (sufficiently
differentiable) congruence of timelike lines, $\gamma(t)$, and for
which the locus of events $t = t_0$ defining the synchronization
constitute a (sufficiently differentiable, transverse) hypersurface,
$\varphi(x) = t_0$. Once the trajectories so synchronized, the loci
of events $t =${\em constant} for {\em any} constant define a
one-parameter family of hypersurfaces, to which the initial
hypersurface $\varphi(x) = t_0$ belongs; let $\varphi(x) = t$ be its
equation.

Any of these hypersurfaces $\varphi(x) = t$  \, is said to define
the {\em same} synchronization that the hypersurface $\varphi(x) =
t_0$. Denoting by $\dot{\gamma}$ the tangent vector to the histories
of the clocks, $\dot{\gamma} \equiv \frac{d}{dt}\gamma(t)$, such
space-time function  $\varphi(x)$
verifies ${\cal L}(\dot{\gamma})\varphi = 1$, where ${\cal L}(\dot{\gamma})$ is
the Lie derivative%
\footnote{On functions $\varphi$ the Lie derivative reduces to a
directional derivative,  ${\cal L}(\dot{\gamma})\varphi$ $=
\dot{\gamma}(d\varphi)$ $= \dot{\gamma}^\rho \partial_\rho\varphi$.}
with respect to $\dot{\gamma}$.

Conversely, it is easy to see that the level hypersurfaces $\psi(x)
= k$, $k =$ {\em constant}, of any function $\psi(x)$ that verifies
${\cal L}(\dot{\gamma})\psi = 1$, define a synchronization for the
(congruence of histories of the) clocks, i.e. there exists a
canonical parameter $t$ for the field $\dot{\gamma}$, $
\frac{d}{dt}\gamma(t) = \dot{\gamma}$, such that $k = t$.

Consequently, for a congruence of (histories of) clocks of tangent
vector field $\dot{\gamma}$, {\em the set of all its possible
synchronizations is the set of all the scalar functions $\psi(x)$
such that ${\cal L}(\dot{\gamma})\psi = 1$}. And it is obvious that,
if $\varphi$ is such a synchronization, {\em any} other
synchronization $\psi$ is of the form  $\psi = \varphi + \omega$,
where $\omega$ is an invariant function of the field $\dot{\gamma}$,
${\cal L}(\dot{\gamma})\omega = 0$. The group of transformations of
(pure) synchronizations for the congruence of clocks, or {\em
synchronization group}, is thus isomorphic to the additive group of
functions $\{\omega\}$ which are invariant for the congruence
$\dot{\gamma}$: if $\varphi$ is an initial synchronization and
$\omega$ any $\dot{\gamma}$-invariant function, any other
synchronization $\psi$ is obtained by $\psi = T_\omega\varphi \equiv
\varphi + \omega$.

To make more explicit the synchronization group as a transformation
group of the space-time, let us start from a coordinate system
$\{x^\alpha\}$ ($\alpha = 0,1, \ldots ,n-1$) adapted both, to the
field $\dot{\gamma}$, say $\dot{\gamma} = \partial_0$, and to the
synchronization $\varphi$, thus $d\varphi = dx^0$. In this
coordinate system, the $\dot{\gamma}$-invariant character of a
function $\omega$ is expressed by its independence of the timelike
coordinate parameter $x^0$, $\omega = \omega(x^i)$, ($i = 1, \ldots,
n-1$). The new coordinate system  $\{X^\alpha\}$, generated by
$\omega$ and adapted both to $\dot{\gamma}$ and to $T_\omega\varphi
= \psi$ is then of the form
\begin{equation}
    \label{grsynchro}
X^0 = x^0 + \omega(x^i) \,\, , \quad X^i = x^i  \,\, .
\end{equation}
{\em These are the space-time transformation equations of the
synchronization group.}

For our purpose here, that of generating easily the Newtonian causal
classes, it is nevertheless sufficient to consider the simplest
subgroup of the synchronization group (\ref{grsynchro}), the {\em
linear synchronization group}:
\begin{equation}
    \label{lingrsynchro}
X^0 = x^0 + a_ix^i \,\, , \quad X^i = x^i  \,\, .
\end{equation}
Its matrix form may be analyzed as follows. Let ${\rm {\bf 1}}$ be
the $n \times 1$ column matrix of components $(1,0
\stackrel{n-1}{\ldots}, 0)$, and consider the set of all the $1
\times n$ matrices ${\bf a}$ orthogonal to ${\rm {\bf 1}}$, ${\bf
a}\cdot{\rm {\bf 1}}=0$; they are obviously of the form $ {\bf a} =
(0,\vec{a})$ with $\vec{a} \equiv (a_1,\ldots,a_{n-1})$. Then, the
{\em linear synchronization algebra} is the (commutative) algebra of
matrices of the form ${\rm {\bf 1}} \otimes {\bf a}$, so that the
matrices $L$ of the linear synchronization group are of the form $L
=$ $exp\{{\rm {\bf 1}} \otimes {\bf a}\} =$ $I + {\rm {\bf 1}}
\otimes {\bf a}$, which clearly correspond to matrices of minimal
polynomial $\left(L - I\right)^2 = 0$. In obvious matrix notation,
equations (\ref{lingrsynchro}) may be written ${\bf X} = L{\bf x}$.

From equations (\ref{lingrsynchro}) we have the relations between
the natural frames and coframes of two coordinate systems related by
a linear synchronization:
\begin{equation}
\label{parci}
\partial_{X^0} = \partial_{x^0}\, \, ,
\qquad  \partial_{X^i} = - a_i
\partial_{x^0} + \partial_{x^i}\, \, ,
\end{equation}
\begin{equation}
\label{diffs}
d X^0 = dx^0 + a_i d x^i \, \, , \qquad d X^i = d x^i\, \, .
\end{equation}

Remark that, until now, all the considerations about the
synchronization group remain valid for both, Newtonian and
relativistic space-times and are applicable to {\em any} starting
coordinate system.

\section{Examples of Newtonian coordinate systems of different causal classes}
\label{sec:5}
\subsection{Generating Newtonian causal classes by the Linear Synchronization Group}
\label{subsec5LinSyncGr}

Surprisingly enough, the linear synchronization group provides one
of the simplest ways of generating {\em all} the Newtonian causal
classes.

In what follows, we will always start, in the Newtonian space-time,
from a {\em standard coordinate system} $\{x^\alpha\}$, that is to
say a coordinate system such that the coordinate lines $x^0 = t,
\,\,  x^i =$ {\em constant} \, are synchronized by the instantaneous
spaces of the absolute time current $\theta$, $dx^0 = \theta = dt$,
and such that the other coordinate lines $x^i =$ {\em variable} are
tangent to these instantaneous spaces, $\gamma^*(\partial_i) = 0$.
Its natural frame is thus of the causal type $\{{\rm t}{\rm
e}\ldots{\rm e}\}$.

Let us apply the transformation (\ref{lingrsynchro}) to this
coordinate system. By construction (definition of a change of
synchronization) the new coordinate $X^0$ is a timelike coordinate
parameter, because $\partial_{X^0}$ is the expression, in this
coordinate system $\{X^\alpha\}$, of $\dot{\gamma}$ , which is
timelike. However, $X^0$ results to be a spacelike gradient
coordinate whenever $\vec{a} \neq 0$, because then, according to
(\ref{diffs}), one has $dX^0 \wedge dt \neq 0$. On the other hand,
every new coordinate $X^i$ is a timelike coordinate parameter
whenever the corresponding component $a_i$ of $\vec{a}$ does not
vanish, because $\partial_{X^i}$, which is given by the second of
expressions (\ref{parci}), is timelike in this case,
$\gamma^*(\partial_{X^i}) \neq 0$. Nevertheless $X^i$ remains a
spacelike gradient coordinate, because $\forall i, dX^i \wedge dt
\neq 0$.

We see thus that, in the $n$-dimensional Newtonian space-time,
starting from a standard  coordinate system $\{t,x^i\}$ of causal
type $\{{\rm t},(n-1){\rm e}\}$, {\em the linear synchronization
transformations {\em (\ref{lingrsynchro})} for every one of the
vectors
$\vec{a}=(1,\stackrel{k-1}{\ldots},1,0,\stackrel{n-k}{\ldots},0)$,
($k = 1, \ldots , n$), define  a coordinate system $\{X^\alpha\}$ of
causal type $\{k{\rm t},(n-k){\rm e}\}$, belonging to the k-th
causal class of the $n$ possible ones}, according to theorem
\ref{teoNwn}. Then, for every $r = 1, \ldots , n$, the ${n \choose
r}$ associated $r$-planes are of causal type $\{[{n \choose r} -
{n-k \choose r}]{\rm T}, {n-k \choose r}{\rm E} \} $.

For $n = 4$, this gives of course the four causal classes of Figure
\ref{New-clases}.

It is worthwhile to note that all the different causal classes have
been obtained by  simple, {\em pure}, changes of synchronization of
the {\em same} system of clocks, excluding any other change of
coordinates or of observers. Apparently, this is not an intuitive
idea for most of us.

\subsection{The causal class of the ancestral local Solar time}
\label{subsec5SolarTime}

The local Solar time, i.e. the time shown by a sundial, is the
oldest timelike coordinate parameter known by humanity, and still
remains indefinitely alive and currently in use, although slightly
deformed by the at present stepped time zones. As we have already
mentioned, this local Solar time is a paradigmatic example of the
situations where the current but particular notion of ``timelike
coordinate'' becomes incoherent.

Specifically, we will consider here the causal class of a coordinate
system at rest with respect to  a spherical Earth in uniform
rotation when the (absolute time rhythmed) clocks are synchronized
by the {\em local Solar time or sundial synchronization}, i.e. are
such that at any place they watch the same fixed  time (say 12h)
when the Sun is just on the local meridian. For simplicity, we have
not taken into account the inclination of the ecliptic and have
neglected the translational motion of the Earth.
\begin{figure}[b]
\centerline{
\parbox[c]{0.67\textwidth}{\includegraphics[width=0.67\textwidth]{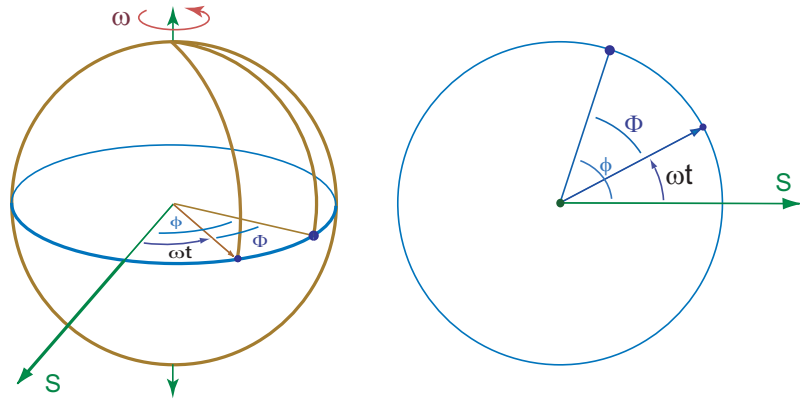}}}
\caption{The geocentric inertial spherical standard coordinates $\{t, r,
   \theta, \phi \}$ and the local Solar time geocentric rotating spherical
   coordinates $\{T, r, \theta, \Phi\}$ are related by
   $T = \frac{\phi}{\omega}, \ \Phi = \phi - \omega t$, where $\omega$ is
   the angular velocity of the Earth. The fixed direction $S$ is that of the
   sun (the inclination of the ecliptic is not taken into account and the
   translational motion of the Earth is neglected). The picture on the right
   shows the Earth equator, $r=R_\oplus,$ $ \theta = 0$, whose history in the
   plane $\{T, \Phi\}$ is represented in Fig. \ref{terra-pla}.
   \label{terra}}
\end{figure}

Let $\{t, r, \theta, \phi \}$ be a standard coordinate system where
$\{r, \theta, \phi \}$ are the usual geocentric inertial spherical
coordinates. This system thus belongs to the standard causal class
$\{{\rm t e e e, T T T E E E}, {\it t e e e}\}$.

The {\em geocentric rotating spherical coordinate system} $\{t, r, \theta,\Phi \}$,
is obviously given by the (pure) rotation
\begin{equation}
\label{Phi-phi}
 \Phi = \phi - \omega t \, ,
\end{equation}
where $\omega$ is the Earth's angular velocity. Here the coordinate
lines where only $t$  varies  are no longer inertial, but the
timelike helices that they describe remain synchronized by the
instantaneous spaces of the time current. This point, and the fact
that the sole new coordinate $\Phi$ verifies $d\Phi \wedge dt \neq
0$, make the causal class of this rotating coordinate system to
remain the standard one.
\begin{figure}
\centerline{
\parbox[c]{0.67\textwidth}{\includegraphics[width=0.67\textwidth]{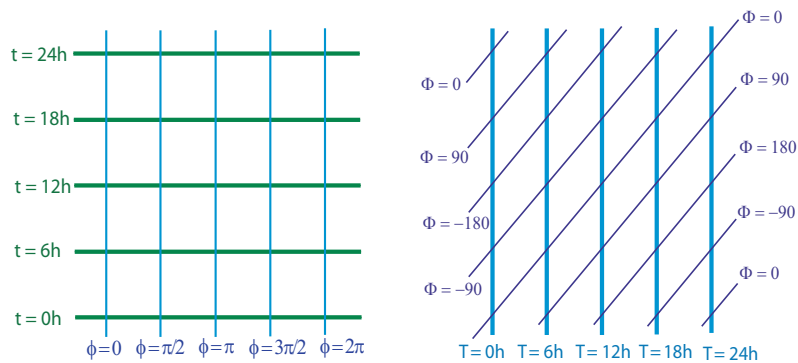}}}
\caption{History of the Earth equator $r=R_\oplus,$ $ \theta = 0$ in the
   plane $\{T, \Phi\}$. (a) In geocentric inertial spherical coordinates:
   the vertical thin straight lines are coordinate lines of the absolute
   time $t$, and the horizontal thick straight lines correspond to the
   absolute synchronization (hypersurfaces of simultaneity $t=constant$).
   (b) In an Earth rotating frame: the histories of the equator events,
   which constitute the coordinate lines of the `solar time' $T$, are
   represented by the inclined thin straight lines $\Phi \equiv \phi-
   \omega t = constant$, meanwhile the `solar synchronization'
   hypersurfaces $T=constant$ are represented by the vertical thick
   straight lines. Note that this `solar instants' contain the coordinate
   lines of the absolute time $t=variable$.
    \label{terra-pla}}
\end{figure}

Now, starting from this coordinate system $\{t, r, \theta,\Phi \}$,
let us perform a (pure) synchronization change of the form
(\ref{lingrsynchro}) to the {\em Solar time geocentric rotating
spherical coordinate system} $\{T, r, \theta,\Phi \}$, that is to
say,
\begin{equation}
\label{TPhi}
 T = t + \frac{\Phi}{\omega} = \frac{\phi}{\omega}\, .
\end{equation}
In this Solar time rotating coordinate system (see Fig.
\ref{terra}), the observers at rest with respect to the rotating
Earth remain at rest by construction, although their local time T
has been synchronized to be given by a sundial for a fixed Sun
placed in the initial system at the meridian of longitude $\phi = 0$
(see Fig. \ref{terra-pla}). So, the coordinate lines where only $T$
varies, of tangent vector $\partial_T$, are timelike.

On the other hand, the coordinate lines where only $\Phi$ varies, of
tangent vector $\partial_\Phi$, are also timelike, because the
inverse transformation is, from (\ref{TPhi}), $\{ t = T -
\Phi/\omega , r, \theta, \phi = \omega T\}$ and it follows $\omega
\partial_\Phi = - \partial_t $: they form, in fact, the congruence
of inertial observers, as it is due. Although we could compute the
causal character of the corresponding $1$-forms, surfaces and
hypersurfaces, this is not necessary because the table of Fig.
\ref{New-clases} gives already this information. Thus, {\em the
causal class of the ancestral local Solar time coordinate system is}
$\{{\rm t t e e, T T T T T E}, {\it e e e e}\}$.

It is worthwhile to note that, in Newtonian physics as well as in
relativity, the more natural and ancestral synchronization is
generated by {\em timelike} hypersurfaces, a fact that seems
systematically forgotten in theoretical physics, where a
synchronization is always defined by {\em spacelike} hypersurfaces.
\subsection{Newtonian Emission Coordinates}
\label{subsec5NewEmission}

Suppose an inertial medium in which a class of signals (sound,
light) propagates at constant velocity $v$. Let $\kappa(t)$ be the
space-time point-like trajectory of an emitter clock that uses such
signals to continuously broadcast his time $t$. In the space-time,
the front waves describe thus (sound-, light-)cones carrying the
value $t =$ {\em constant}. Four such emitters $\kappa^A(t)$  ($A =
1,2,3,4$) fill the space-time with four (one-parameter) families of
cones $t^A =$ {\em constant} which generically define a space-time
system of {\em emission coordinates}.

Let us take every event as the vertex of the past cone of the
velocity $v$ of propagation of the class of signals in question.
This cone cuts the four histories $\kappa^A(t)$ of the clocks at the
clock times $t^A$. Then, the set $\{t^A\}$ constitutes the four
emission coordinates of the event.

Here we will consider the simple case of four emitters at rest with
respect to the inertial medium referred to a standard coordinate
system $\{t,x^i\} = \{t,\vec{r}\}$, of worldlines
\begin{equation}
\label{EqEmit}
 \kappa^A(t) = (t,\vec{c}^A)\, .
\end{equation}
Then, the signal emitted by the clock $\kappa^A$ at the instant
$t^A$ at velocity $v$ describes in the space-time a cone of equation
\begin{equation}
\label{EqCon}
 v(t-t^A) = \left|\vec{r}-\vec{c}^A\right|\, ,
\end{equation}
so that the emission coordinates $\{t^A\}$ are related to the inertial ones
$\{t,\vec{r}\}$ by
\begin{equation}
\label{tAtr}
 t^A = t - \frac{1}{v}\left|\vec{r}-\vec{c}^A\right|\, .
\end{equation}

To know the causal class of the emission coordinates $\{t^A\}$ it is
convenient to consider the coordinate $r$-forms. From (\ref{tAtr}),
the coframe of $1$-forms $\{dt^A\}$ may be written
\begin{equation}
\label{Oneforms}
 dt^A = dt + \omega^A \,, \quad    \omega^A \equiv - \frac{1}{v} u^A \, ,
\end{equation}
where $u^A$ is the $1$-form associated to the generically unit
spacelike vector $\vec{u}^A$,  given by
\begin{equation}
\label{uA}
 \vec{u}^A \equiv \frac{\vec{r}-\vec{c}^A}{\left|\vec{r}-\vec{c}^A\right|} \, ,
\end{equation}
$u^A = \gamma(\vec{u}^A)$, $\gamma$ being the $3$-dimensional
inverse of the structure metric $\gamma^*$ associated to the
inertial observers $\partial_t$\,,  $\gamma.\gamma^* = I - \theta
\otimes \partial_t$\,, and $\theta$ being the time current%
\footnote{Note that, meanwhile $\gamma^*$ is an intrinsic element of
the geometry of Newtonian space-time, its `three-dimensional
inverse' $\gamma$ is an {\em  observer-dependent} quantity, given by
$\gamma.\gamma^* = I - \theta \otimes u$, where $u$ is the unit
velocity of the chosen observer. To two different observers, they
correspond two different degenerate four-dimensional covariant
metrics $\gamma$ of rank three, although their induced spatial
components on the instantaneous space take the same value, as it is
well experienced in the usual three-dimensional formalism.}.
The Jacobian matrix of the transformation (\ref{tAtr}) is not
defined at the events $(t,\vec{r})$ where $\vec{r} = \vec{c}^A$\,,
that is to say, along the clock worldlines $\kappa^A$. Below we
shall see other events where the Jacobian matrix is not defined. Out
of these worldlines one has $\omega^A \neq 0$ and thus $dt^A$ is
spacelike (it is not collinear to the time current). Consequently,
{\em the coframe of the Newtonian emission coordinate system is of
causal type $\{e\,e\,e\,e\}$.}
\begin{figure}
\centerline{\parbox[c]{0.65\textwidth}
{\includegraphics[width=0.50\textwidth]{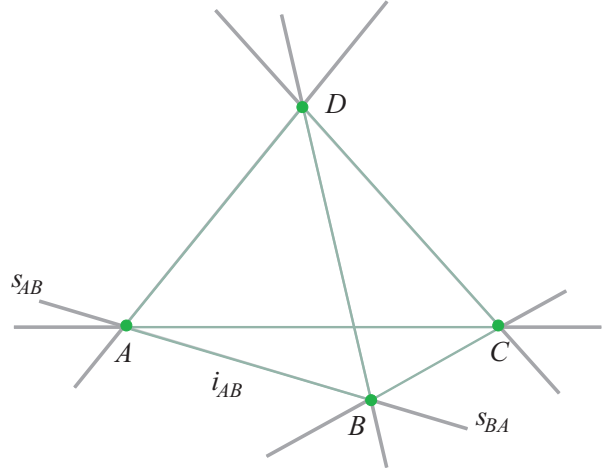}}}
\caption{At any instant $t =$ {\em constant}, the positions
$\kappa^A(t) \equiv A$ ($A = 1,2,3,4$) of the four clocks
generically define the four vertices $A,B,C,D$ (all $\neq$) of a
$3$-dimensional tetrahedron. If the clocks are at rest in an
inertial system, the outer open segments $s_{AB}$ and $s_{BA}$ of
the straight line $\ell_{AB}$ containing the edge  $i_{AB}$ between
the vertices $A$ and $B$ represent the shadows of the signals $B$
and $A$ respectively produced by $A$ and $B$. \label{fig-tetra}}
\end{figure}

The co-planes of the coordinate system are determined by the $2$-forms
\begin{equation}
\label{coplanes} dt^A \wedge dt^B = dt \wedge (\omega^B - \omega^A )
+ \omega^A \wedge \omega^B \, ,
\end{equation}
so that the co-plane $AB$ is generically spacelike, and can be
timelike only when $\omega^A \wedge \omega^B = 0$\,,  that is to say
on the timelike plane of events $\Pi_{AB}$ that contains the
worldlines of $\kappa^A$ and $\kappa^B$. Because the clocks are at
rest with respect to the starting inertial system, at any $t =$ {\em
constant} their positions $\kappa^A(t) \equiv A$  will generically
define the four vertices $A,B,C,D$ (all $\neq $) of a
$3$-dimensional tetrahedron (see Figure \ref{fig-tetra}). Denote by
$\ell_{AB}$ the straight line passing through $A$ and $B$ and, in
it, by $i_{AB} $ the corresponding open edge of the tetrahedron and
by $s_{AB}$ ({\em respect.} $s_{BA}$) the other open segment
contiguous to $A$ ({\em respect.} contiguous to $B$). It is then
clear that the timelike plane $\Pi_{AB}$ is the history of the
straight line $\ell_{AB}$, and we will denote by $I_{AB}$ the
history of $i_{AB} $, the (timelike) open strip of $\Pi_{AB}$ whose
boundaries are $\kappa^A$ and $\kappa^B$. Similarly, $S_{AB}$ ({\em
respect.} $S_{BA}$) will denote the (timelike) open strip of
$\Pi_{AB}$ contiguous to $\kappa^A$ ({\em respect.} contiguous to
$\kappa^B$). Now we see that the condition $\omega^A \wedge \omega^B
= 0$\, takes place along $\ell_{AB}$,  thus on the events of
$\Pi_{AB}$. In addition, because from (\ref{Oneforms}) all the
$\omega^A$ have same length, one has $\omega^A = - \omega^B$ on
$i_{AB}$, thus on the events of $I_{AB}$, and $\omega^A = \omega^B$
on the two other open segments $s_{AB}$ and $s_{BA}$, thus  of the
events of $S_{AB}$ and $S_{BA}$, where one has
\begin{equation}
\label{dts}
dt^A \wedge dt^B = 0 \, ,
\end{equation}
and the coordinate system degenerates. These open strips  of
$\Pi_{AB}$, $S_{AB}$ and $S_{BA}$, are also the half-planes
describing the history of the {\em shadows} that the clocks $A$ and
$B$ make respectively to the signals of the  clocks $B$ and $A$.
These considerations on expressions  (\ref{coplanes}) and
(\ref{dts}) show that either all the coordinate coplanes are
spacelike, or one of them is  timelike, so that, on account of lemma
\ref{lem3}, it results that {\em generically the type of the
coordinate planes is} $\{{\rm T\,T\,T\,T\,T\,T}\}$ {\em but on the
events of the six timelike strips} $I_{AB}$, {\em and only on them,
the type is} $\{{\rm T\,T\,T\,T\,T\,E}\}$, {\em the coordinate
system being degenerate on the shadows} $S_{AB}$ {\em and} $S_{BA}$
{\em and undetermined on the worldlines} $\kappa^A$.

To analyze the coordinate lines, let us consider the dual $3$-forms:
\begin{equation}
  \begin{array}{lcl}
  \label{3forms}
  dt^A \wedge dt^B \wedge dt^C &=& \omega^A \wedge \omega^B \wedge \omega^C  \\
 &+& dt \wedge \left(\omega^A \wedge \omega^B + \omega^B \wedge \omega^C + \omega^C
\wedge \omega^A \right)  \, .
  \end{array}
\end{equation}
The $3$-coplane  $ABC$ is generically spacelike, and can be timelike
only when $\omega^A \wedge \omega^B \wedge \omega^C = 0$, what
happens on the events of the timelike $3$-plane $\Pi_{ABC}$ that
contains the worldlines $\kappa^A,$ $\kappa^B,$ $\kappa^C$. In the
stationary $3$-dimensional sections $t =$ {\em constant}, these
events correspond to the planes $\ell_{ABC}$ that contain the three
clocks $A$, $B$, $C$, and thus the three lines $\ell_{AB}$,
$\ell_{BC}$, $\ell_{CA}$, including the tetrahedral faces $i_{ABC}$
that their edges $i_{AB}$, $i_{BC}$ and $i_{CA}$ delimit, and the
six strips $s_{AB}$, $s_{BA}$, $s_{BC}$, $s_{CB}$, $s_{CA}$,
$s_{AC}$. We already know that, apart from on the clocks $A$, $B$,
$C$ themselves,  on these last six strips  the coordinate coplanes
degenerate; are there other events than that on which the coordinate
$3$-coplanes be degenerate? In other words, there where  $\omega^A
\wedge \omega^B \wedge \omega^C = 0$ out of the edges, can the other
term in (\ref{3forms}) also vanish? We have:
\begin{equation}
    \label{omegaslindp}
    \omega^C = \alpha \omega^A + \beta \omega^B  \, ,
\end{equation}
so that (\ref{3forms}) becomes
\begin{equation}
  \label{dege3forms}
   dt^A \wedge dt^B \wedge dt^C = (1 - \alpha - \beta) dt \wedge \omega^A
\wedge   \omega^B   \, ,
\end{equation}
which cannot degenerate, being $\omega^A \wedge\omega^B \neq 0$,
unless $\alpha + \beta = 1$. But
\begin{equation}
  \label{omegac}
   1 = (\omega^C)^2 = \alpha^2 + \beta^2 + 2\alpha\beta(\omega^A\cdot\omega^B) =
   1 + 2\alpha\beta(\omega^A\cdot\omega^B - 1)\, ,
\end{equation}
admits no solution, because $\alpha \neq 0 \neq \beta$ and
necessarily $\omega^A\cdot\omega^B < 1$. The tangent vectors to the
coordinate lines being at every event causally related to the
$3$-planes by lemma \ref{lem3}, it results the following.

{\em The coordinate lines of the emission coordinates in Newtonian
space-times are generically of type} $\{{\rm t\,t\,t\,t}\}$, {\em
but on the events of the timelike $3$-planes} $ \Pi_{ABC}$ {\em
containing three emitters they are generically  of type} $\{{\rm
t\,t\,t\,e}\}$, {\em and are of type} $\{{\rm t\,t\,e\,e}\}$ {\em on
the events of the timelike strips} $I_{AB}$ {\em generated by every
pair of clocks.}

It is pertinent here to note that, {\em  in Newtonian space-time,
the emission coordinate system generated by a positioning system is
never causally homogeneous}, but always presents three regions
corresponding to the non standard three causal classes. Only the
emission coordinate systems generated by  relativistic positioning
systems based in light signals are always causally homogeneous, as
we will see in next section.

The geometry of the coordinate surfaces and coordinate lines of the
emission coordinates is simple. Because generated by the  two by two
or three by three intersections of the coordinate hypersurfaces,
which are isotropic cones of parallel axes,  {\em the coordinate
surfaces and coordinate lines of the emission coordinates are
hyperboloids and hyperbolas respectively.} As already seen, these
hyperbolas are generically timelike lines, up to at their base
point, where they become spacelike.

As we have seen, the transformation (\ref{tAtr}) from a standard
inertial coordinate system $\{t,x^i\} = \{t,\vec{r}\}$ to an
emission coordinate system $\{t^A\}$ is degenerate on the clock
shadows $S_{AB}$, timelike space-time surfaces generated by every
clock for the signals coming to it from the others. Thus the
question: is transformation (\ref{tAtr}) degenerate at other events
than those of the shadows $S_{AB}$? To see it, let us consider the
coordinate volume ${\rm element}\,\, \eta_{ec}$:
\begin{equation}
  \begin{array}{lcl}
  \label{4forms}
  \eta_{ec} & \equiv & dt^A \wedge dt^B \wedge dt^C \wedge dt^D \\
            &=& dt \wedge \left[ \right. \begin{array}[t]{ll}
            \!\!\! -& \!\!\! \omega^A
            \wedge \omega^B \wedge \omega^C +
            \omega^B \wedge \omega^C \wedge \omega^D  \\
            \!\!\!  -& \!\!\! \omega^C \wedge \omega^D
            \wedge \omega^A + \omega^D \wedge \omega^A \wedge
            \omega^B \left. \right]
            \end{array}  \\
            &=& - dt \wedge \left[ (\omega^A - \omega^D) \wedge
            (\omega^B - \omega^D) \wedge (\omega^C - \omega^D) \right]  \, .
            \end{array}
\end{equation}
It is then evident that the Jacobian is degenerate, as we already
know, there where $\omega^A = \omega^B$, that is to say, on the
clock shadows $S_{AB}$, for any pair $A\neq B$. But (\ref{4forms})
shows that it can be also degenerate there where the three vectors
$\omega^A - \omega^D$ are linearly dependent. It can be seen (for
example in \cite{Escola}), that this happens on the events where the
signals coming from the four clocks are seen or heard as coming from
four points located on a circle of the celestial sphere of the event
(quotient of the instantaneous space of the event by the radial
distance to the event).

\section{Lorentzian causal classes with Newtonian
analogues}
\label{sec:6}

Theorem \ref{teo-199} establishes the existence of $199$ Lorentzian
causal classes of space-time frames \cite{199}.  Among them, one can
found the analogue to the $4$  Newtonian causal classes of
space-time frames \cite{Escola} i.e., four Lorentzian classes of
frames having the same causal signature that the four Newtonian
ones.  Thus, whatever be the relativistic space-time, one can always
choose local coordinate systems belonging, in some region, to any of
the $199$ causal classes and, in particular, having the same causal
signature that any given Newtonian coordinate system. But it must be
emphasized that going from $4$ to $199$ causal classes, the change
from the Newtonian conception of the space-time to the relativistic
one implies a richness of causally different ways of locating
space-time events that, in spite of the appearances extracted from
the current scientific publications, is far from being well
understood.

Here, we shall analyze in Minkowski space-time the situations that
we have already analyzed in the Newtonian case.

\subsection{The Linear Synchronization Group}
\label{LinSynchroMinko}

Let us consider, in Minkowski space-time, the linear synchronization
group (\ref{lingrsynchro}) acting on an inertial laboratory referred
to a standard coordinate system $\{x^0, x^i\}$. The metric
components $\eta_{\alpha \beta}$ of the Minkowski flat metric $\eta$
in this coordinate system are the usual $\eta_{\alpha \beta} = {\rm
diag}(-1,1, \ldots ,1)$ so that the associated natural frame is of
the causal type $\{{\rm t} \, {\rm e}\ldots{\rm e}\}$.

The natural frame and coframe of the new system $\{X^{\alpha}\}$ are
given by (\ref{parci}) and (\ref{diffs}). It follows, by direct
scalar products of these expressions, that the covariant and
contravariant components, $g_{\alpha \beta}$ and $g^{\alpha \beta}$
respectively, of the metric $\eta$ in this new system are:
\begin{equation} \label{metric-linear}
g_{\alpha \beta} = \left(
\begin{array}{cc}
-1  & \vec{a} \\
\vec{a}  & I - \vec{a} \otimes \vec{a} \\
\end{array} \right) \, , \hspace{1cm} g^{\alpha \beta} = \left(
\begin{array}{cc}
-1 + \vec{a}\,^2 & \vec{a} \\
\vec{a}  & I \\
\end{array} \right) \, .
\end{equation}
where $\vec{a} \equiv (a_1,\ldots,a_{n-1})$, $\vec{a}\,^2 \equiv
\sum_{i=1}^{n-1} a_i^2$ and $I$ is the
$n\hspace{-1mm}-\hspace{-1mm}1$ identity matrix.

We can see from (\ref{metric-linear}) that, {\em like in the
Newtonian case, the new coordinate $X^0$ is a timelike coordinate
parameter}. However, {\em $X^0$ results to be a spacelike gradient
coordinate only when $| \vec{a} | > 1$, meanwhile in the Newtonian
case the condition is simply  $\vec{a} \neq 0$. When $| \vec{a} | =
1$ or $| \vec{a} |< 1$, $X^0$ is a null or timelike gradient
coordinate, respectively}. Obviously, the first of these last two
situations is forbidden in the Newtonian case, and the second one
cannot be attained by the linear synchronization group (up to,
trivially, by the identity transformation, $ \vec{a}  = 0$).

On the other hand, {\em every new coordinate $X^i$ remains, like in
the Newtonian case, a spacelike gradient coordinate}. However {\em
$X^i$ results to be a timelike  coordinate parameter only when $|
a_i |> 1$, meanwhile in the Newtonian case the condition is simply
$a_i \neq 0$. When $| a_i | = 1$ or $| a_i | < 1$, $X^i$ is a null
or spacelike coordinate parameter, respectively}. Both situations
are also absent in the Newtonian case (up to for $ \vec{a}  = 0$).

Finally, the coordinate two-forms satisfy:
\begin{eqnarray}
dX^i \wedge  dX^j = dx^i \wedge  dx^j\; , & \quad dX^0 \wedge dX^i = dx^0
\wedge dx^i + a_j dx^i \wedge dx^j \; , \\
(dX^i \wedge dX^j)^2 = 1 \;, & \quad (dX^0 \wedge dX^i)^2 = -1 +
\vec{a}\,^2 - a_i^2\; .
\end{eqnarray}
Consequently, the $(n-2)$-coordinate surfaces $X^i = constant$, $X^j
= constant$ ($i,j$ given) are timelike and the $(n-2)$-coordinate
surfaces $X^0 = constant$, $X^i = constant$ ($i$ given) are
timelike, null or spacelike if  $\vec{a}\,^2 - a_i^2$ is greater,
equal or smaller than $1$, respectively. This information,
insufficient for $n > 4$, completely determines the causal class of
the coordinate system $\{X^0, X^i\}$ in the four-dimensional
Minkowski space-time:

{\em All  the causal classes  obtained by a linear synchronization
transformations have a causal signature of the form:
\begin{equation}
\{{\rm t \, c_1 \, c_2 \, c_3, \, T \, T \, T \, C_{12} \, C_{13} \,
C_{23}}, \, c_0 \, e \, e \, e \}
\end{equation}
where the non-fixed causal orientations, ${\rm c_1, c_2, c_3, C_{12},
C_{13}, C_{23}}, c_0$ depend on the $a_i$ parameters as follows:}
\begin{equation}
\hspace{-2.0cm}
{\rm c_i} = \cases{{\rm t} \qquad  | a_i | >1 \cr {\rm l} \qquad |
a_i | = 1 \cr {\rm e} \qquad  | a_i | < 1  } \qquad
{\rm C_{ij}} = \cases{{\rm T} \qquad   a_i^2 + a_j^2 > 1 \cr {\rm L}
\qquad  a_i^2 + a_j^2  = 1 \cr {\rm E} \qquad   a_i^2 + a_j^2 < 1 }
\qquad
c_0 = \cases{t \qquad  | \vec{a} | < 1 \cr l \qquad  | \vec{a} | = 1
\cr e \qquad  | \vec{a} | > 1  }
\end{equation}

A more detailed analysis of the compatible orientations shows that
{\em the number of different causal classes that may be generated by a
linear synchronization transformation is $29$, in contrast with
the only $4$ Newtonian ones}.
We will consider them elsewhere \cite{lorentziano}.

Evidently the four Newtonian analogues exist in relativity. In fact,
{\em the Lorentzian causal classes of same causal signature that the
four Newtonian ones
 correspond to the following values of the parameters $a_i$:}
\begin{equation*}
\begin{array}{lllr}
\{{\rm t \, t \, t \, t, \, T \, T \, T \, T \, T \, T}, \, e \, e
\, e \, e \}\  & {\rm if} & \ \ \  \forall \ i, & |a_i| > 1 \ \, \\[2mm]
\{{\rm t \, t \, t \, e, \, T \, T \, T \, T \, T \,
    T}, \, e \, e \, e \, e \}\  & {\rm if} & \cases{\exists ! \ i, \\
                                                   \forall \ j \not = i,} &
                                                  \begin{array}{r}
                                                    |a_i| < 1 \\
                                                    |a_j| > 1
                                                  \end{array}\\[6mm]
\{{\rm t \, t \, e \, e, \, T \, T \, T \, T \, T \, E}, \, e \, e
\, e \, e \}\   & {\rm if}  & \cases{\exists ! \ i, \\
                                     j,k \not=i,} &
                                    \begin{array}{r}
                                    |a_i| > 1 \\
                                    a_j^2+a_k^2 < 1
                                    \end{array} \\[6mm]
\{{\rm t \, e \, e \, e, \, T \, T \, T \, E \, E \,  E}, \, t \, e
\, e \, e \}\  & {\rm if} & \ \ \ \forall\ i, & |a_i| < 1 \ \,
\end{array}
\end{equation*}

\vspace{3mm}

\subsection{The local Solar time synchronization.}
\label{rotation}

In the Newtonian example of the rotating Earth of subsection
\ref{subsec5SolarTime}, the latitude of the observer plays in fact
no role, because we are interested only in the time synchronization,
not in the angular height of the Sun. For this reason, we shall
consider here, in place of the Earth, a rigidly rotating disk and,
in place of spherical coordinates, cylindrical ones.

So, let $\{t, \phi, \rho, z \}$ be an inertial laboratory referred
to a standard cylindrical coordinate system in Minkowski space-time.
This coordinate system is known to belong to the standard causal
class $\{{\rm t e e e, T T T E E E}, {\it t e e e}\}$.

The {\em rotating cylindrical coordinate system} $\{t, \Phi, \rho, z \}$,
adapted to the congruences of the observers in rigid rotation motion
is defined by the transformation (\ref{Phi-phi}). In the Newtonian case
this system remains in the standard class, as happens for the rotating
spherical coordinate system considered in subsection
\ref{subsec5SolarTime}.

As it is well known, in Minkowski space-time the light cylinder
$\rho = 1/\omega$ generates other causal classes. Indeed, the
covariant and contravariant components of the metric tensor in this
rotating coordinate system are, respectively:
\begin{equation} \label{metric-rigid}
\hspace{-1.2cm} g_{\mu \nu} = \left(
\begin{array}{cccc}
-1 + \omega^2 \rho^2  & \omega \rho^2 & \, 0 \, & \, 0  \,\\
\omega \rho^2 & \rho^2 & 0 & 0\\
0 & 0 & 1 & 0 \\
0 & 0 & 0 & 1
\end{array} \right) , \quad g^{\mu \nu} = \left(
\begin{array}{cccc}
-1  & \omega  & \, 0 \, & \, 0  \,\\
\omega  & \frac{1}{\rho^2} - \omega^2 & 0 & 0\\
0 & 0 & 1 & 0 \\
0 & 0 & 0 & 1
\end{array} \right) .
\end{equation}

From here we easily obtain the following causal classes:
\begin{eqnarray}
\{{\rm t \, e \, e \, e, \, T \, T \, T \, E \, E \, E}, \, t \, e
\, e \, e \} \qquad & {\rm if} \quad \rho < 1/\omega  \label{rhomenor}  \\
\{{\rm l \, e \, e \, e, \, T \, L \, L \, E \, E \, E}, \, t \, l
\, e \, e \} \qquad & {\rm if} \quad \rho = 1/\omega\\
\{{\rm e \, e \, e \, e, \, T \, E \, E \, E \, E \, E}, \, t \, t\,
e \, e \} \qquad & {\rm if} \quad \rho > 1/\omega
\end{eqnarray}

The causal orientation ${\rm c_{\alpha}}$ of the vectors of the
coordinate frame is given by the sign of the diagonal elements
$g_{\alpha\alpha}$ of the metric matrix; correspondingly, the causal
orientation ${\rm C_{\alpha\beta}}$ of the coordinate 2-surfaces
is given by the signs of the second order diagonal minors,
$g_{\alpha\alpha} g_{\beta\beta}- (g_{\alpha\beta})^2$; and finally
the causal orientation $c_\alpha$ of the coordinate co-frame is given
by the signs of the diagonal elements $g^{\alpha\alpha}$
of the inverse metric matrix $g^{\mu\nu}$.

Note that, {\em in the rotating system, $t$ remains a timelike gradient
coordinate},  which determines the events that are simultaneous with respect to the
inertial observer at rest at the rotation axis. Nevertheless, $t$ {\em  only remains
a timelike coordinate parameter in the interior of the light cylinder,} $\rho <
1/\omega$.

The timelike helices $t =$ {\em variable} are thus synchronized with an
inertial time. But in the region $\rho \geq 1/\omega$ they become null or
spacelike helices and they do not represent the history of a system of
observers in rigid motion, as it is well known.

Now, starting from this rotating system  $\{t, \Phi, \rho, z \}$,
let us perform the Solar time linear synchronization change
(\ref{TPhi}). In the new coordinate system $\{T, \Phi, r, \theta
\}$, the covariant and contravariant components of the metric tensor
are, respectively:
\begin{equation} \label{metric-solartime}
\hspace{-1.2cm} g_{\mu \nu} = \left(
\begin{array}{cccc}
-1 + \omega^2 \rho^2  &  \frac{1}{\omega} & \, 0 \, & \, 0  \,\\
\frac{1}{\omega} &  -\frac{1}{\omega^2} & 0 & 0\\
0 & 0 & 1 & 0 \\
0 & 0 & 0 & 1
\end{array} \right) , \quad g^{\mu \nu} = \left(
\begin{array}{cccc}
\frac{1}{\omega^2 \rho^2}  &   \frac{1}{\omega \rho^2}  & \, 0 \, & \, 0  \,\\
\frac{1}{\omega \rho^2} & \frac{1}{\rho^2} - \omega^2 & 0 & 0\\
0 & 0 & 1 & 0 \\
0 & 0 & 0 & 1
\end{array} \right) .
\end{equation}

From these coefficients, it then follows that, {\em in the interior
$\rho < 1/\omega$ of the light cylinder, the Solar time rotating
coordinate  system $\{T,\Phi, r, \theta \}$  belongs to the causal class
$\{{\rm t t e e, T T T T T E}, {\it e e e e}\}$, of same causal signature that the  Solar time geocentric rotating system of Newtonian space-time}.

On the light cylinder $\rho=1/\omega$ the new coordinates belong to the causal class $\{{\rm t l e e, T T T L L E}, {\it l e e e}\}$ and on the exterior region  $\rho>1/\omega$ it becomes the standard class $\{{\rm t e e e, T T T E E E}, {\it t e e e}\}$.

\subsection{Relativistic Emission Coordinates}
\label{subsec4RelEmission}

Let us consider now the relativistic analog of the emission
coordinates defined in subsection \ref{subsec5NewEmission}. Now,
every emitter $\kappa$ is supposed to continuously broadcast, in an
inertial non-dispersive medium, their proper time $\tau^A$ by means
of sound or light signals that propagate in the medium at constant
velocity $v \leq 1$.

As in subsection \ref{subsec5NewEmission}, the four emitters will be
consider at rest with respect to the medium referred to a standard
coordinate system $\{t,x^i\} = \{t,\vec{r}\}$. Then, the inertial
time $t$ is also the proper time of the four emitters and their
worldlines take the expression (\ref{EqEmit}): $\kappa^A(t) =
(t,\vec{c}^A)\,$. Then, the equation of the cones that describe the
signals is (\ref{EqCon}), and the emission coordinates $\{t^A\}$ are
related to the inertial ones $\{t,\vec{r}\}$ by (\ref{tAtr}).

Let us first consider the (sound) case $v < 1$.

To know the causal class of the emission coordinate system $\{t^A\}$ we
can start from the coframe of $1$-forms $\{dt^A\}$ given in
(\ref{Oneforms}) and (\ref{uA}). Out of the clock worldlines
$\kappa^A$, where the transformation (\ref{tAtr}) is not defined,
$dt^A$ is spacelike because:
\begin{equation}
\label{dtdos}
(dt^A)^2 = -1 + \frac{1}{v^2} > 0
\end{equation}
Consequently, {\em the coframe of the relativistic emission coordinate
system with $v < 1$ is of causal type} $\{e\,e\,e\,e\}$.

The co-planes of the coordinate system are determined by the
$2$-forms (\ref{coplanes}) that satisfy
\begin{equation} \label{tAtB}
(dt^A \wedge dt^B)^2 = - \frac{1}{v^4} (\mu_{AB}^2 - 2 v^2 \mu_{AB}
+2 v^2 -1) \, , \quad  \mu_{AB} \equiv u_{A} \cdot u_{B} \, .
\end{equation}
Note that $\mu_{AB}$ is the cosine of the angle between the signals
coming from the emitters $A$ and $B$. The study of the polynomial (\ref{tAtB}) in $\mu_{AB}$  leads to the following: {\em the co-plane $AB$ is
spacelike, null or timelike according as $\mu_{AB}$ is greater,
equal or smaller than} $2v^2-1$.

To analyze the coordinate lines, let us consider the dual $3$-forms
(\ref{3forms}). We have:
\begin{equation}
   \label{tAtBtC}
 (dt^A \wedge dt^B \wedge dt^C)^2 = \frac{1}{v^4}
 \left(\frac{1-v^2}{v^2} \Delta_D - \Lambda_D\right) \, , \quad D
 \not= A,B,C \, ,
\end{equation}
where $\Delta_D$ and $\Lambda_D$ depend on $\mu_{AB}$ as:
\begin{eqnarray}
   \label{Delta}
     \Delta_D \equiv (u_{A} \wedge u_{B} \wedge u_{C})^2 = 1 + 2
   \mu_{AB} \mu_{BC} \mu_{CA} - (\mu_{AB}^2 + \mu_{BC}^2 +
   \mu_{CA}^2)\\ \label{Lambda}
   \Lambda_D \equiv 2 (1 - \mu_{AB})(1- \mu_{BC})(1- \mu_{CA})
\end{eqnarray}
Thus, {\em the $3$-coplane  $ABC$ is spacelike, null or timelike
according as $\frac{\Lambda_D}{\Delta_D}$ is smaller, equal or
greater than} $\frac{1-v^2}{v^2}$.

From this information it follows that {\em the causal classes  of the
emission coordinate systems $\{t^A\}$ are of the form:}
\begin{equation}
\{{\rm c_1 \, c_2 \, c_3 \, c_4, \, C_{12} \, C_{13} \, C_{14} \,
C_{23} \, C_{24} \, C_{34}} , \, e \, e \, e \, e \}
\end{equation}
{\em where the causal orientations, ${\rm c_A}$, ${\rm C_{AB}}$
depend on the cosines $\mu_{AB}$ of the angles between the signals
coming from the emitters $A$ and $B$ as:}
\begin{equation} \label{causal-types-e-r}
\hspace{-1.5cm}
{\rm c_A} = \cases{{\rm t} \qquad  \frac{\Lambda_A}{\Delta_A} <
\frac{1-v^2}{v^2}  \cr {\rm l} \qquad \frac{\Lambda_A}{\Delta_A} =
\frac{1-v^2}{v^2} \cr {\rm e} \qquad \frac{\Lambda_A}{\Delta_A} >
\frac{1-v^2}{v^2}  } \qquad \qquad
{\rm C_{AB}} = \cases{{\rm T} \qquad   \mu_{CD} > 2v^2-1 \cr {\rm L}
\qquad  \mu_{CD}  = 2v^2-1 \cr {\rm E} \qquad \mu_{CD} < 2v^2-1 }
\end{equation}
{\em with} $C,D \not= A,B$. A more detailed analysis, which will be
presented elsewhere \cite{lorentziano},  of the compatible
orientations lead to the following result: {\em depending on the
different configurations of the stationary emitters and/or of the
different values of the velocity} $v < 1${\em , the emission
coordinate systems may present space-time regions of 102 different
causal classes.}

It is worth mentioning that, {\em some emitters' configurations and
sound velocities $v < 1$, generate space-time regions of the same
causal signatures that those of the three Newtonian cases}
(subsection \ref{subsec5NewEmission}).

Indeed, {\em the three Newtonian causal signatures are related to
how the events receive the sound signals, according to the following
three sets of conditions:}

\begin{equation*}
\begin{array}{llr}
\{{\rm t \, t \, t \, t, \, T \, T \, T \, T \, T \, T}, \, e \, e
\, e \, e \}\ & {\rm if} & \qquad \ \ \forall A$, $\ \, \qquad \,
\displaystyle \frac{\Lambda_A}{\Delta_A} < \frac{1-v^2}{v^2} \\[6mm]
\{{\rm t \, t \, t \, e, \, T \, T \, T \, T \, T \, T}, \, e \, e
\, e \, e \}\ & {\rm if} &  \displaystyle \cases{ \ \ \ \exists ! \
A, \   \qquad \, \frac{\Lambda_A}{\Delta_A} > \frac{1-v^2}{v^2} \cr
\ \displaystyle \forall B \not = A,  \ \  \quad \, \displaystyle
\frac{\Lambda_B}{\Delta_B} < \frac{1-v^2}{v^2}}\\[12mm]
\{{\rm t \, t \, e \, e, \, T \, T \, T \, T \, T \, E}, \, e \, e
\, e \, e \}\ & {\rm if} & \displaystyle \cases{{\rm for}\ I=A,B, \,
\, \frac{\Lambda_I}{\Delta_I} < \frac{1-v^2}{v^2} \cr \, \forall C
\not = A,B,  \ \ \, \frac{\Lambda_C}{\Delta_C} > \frac{1-v^2}{v^2}
\cr \displaystyle
\mu_{AB} < 2v^2-1} \\[6mm]
\end{array}
\end{equation*}

\vspace{3mm}

Finally, let us consider the (light) case $v=1$. In this case it is
clear that, unlike (\ref{dtdos}), we have $(dt^A)^2 = 0$ so that
{\em the coframe of the relativistic emission coordinate systems
with $v = 1$ is of causal type} $\{l \, l \, l \, l \}$. It can be
then shown that the other causal orientations ${\rm c_A}$ and ${\rm
C_{AB}}$ are  recovered by making $v=1$ in (\ref{causal-types-e-r}).
From  expressions (\ref{causal-types-e-r}), because $\Lambda_A$ and
$\Delta_A$ are both positive and the $\mu_{AB}$ are all smaller than
$1$,  the second members of the expressions for ${\rm c}_A$ vanish
and   those for the ${\rm C}_{AB}$ take the value $1$, the $c_A$ and
the ${\rm C}_{AB}$ cannot but be space-like, ${\rm c}_A = {\rm e}$,
${\rm C}_{AB} = {\rm E}.$  This result, obtained for an inertial
homogeneous medium and four static clocks, may be shown true also
for arbitrary clocks in general space-times \cite{4D}. We have thus:
{\em all the relativistic positioning systems with light signals
define in their whole domains a sole causal class, of causal
signature}
  $$\{{\rm e \, e \, e \, e \, , E \, E \, E \,
E \, E \, E} , \, l \, l \, l \, l \}$$
These relativistic positioning systems, of great interest for future
space research and navigation, have been considered elsewhere
\cite{2D-A,2D-B,4D}.

For same reasons that in the Newtonian case, the coordinate lines of
emission coordinates are here also hyperbolas. Nevertheless, their
causal types differ: meanwhile in the Newtonian case every hyperbola
is everywhere time-like up to at its base point, where it is
space-like, in the relativistic case with $v < 1$ the corresponding
space-like point becomes  enlarged to a whole space-like domain,
bounded by two light-like points, the rest of the branches being
time-like. In the relativistic case $v = 1$ the hyperbolas are
spacelike everywhere. Obviously, this is at the basis of the
richness (the above  mentioned $103$ causal classes) of the
relativistic positioning systems.

\section{Comments around our results}
\label{sec:7}

That the causal structure of the relativistic spacetime allows to
locally classify coordinates systems in $199$ causal classes is
known from some time ago \cite{199}. Nevertheless, the corresponding
situation for Newtonian space-time has remained unanswered. We have
here solve it, showing that in Newtonian space-time the number of
causal classes of coordinate systems reduces to only $4$ (theorem
\ref{teoNw}).

Of these four classes, the standard one, of causal signature $\{{\rm
t \, e \, e \, e,}$ $ \,{\rm T \, T \, T \, E \, E \, E} ,$ $ \, t
\, e \, e \, e \}$, seems to be the only class of which many people
is aware or, at least, the only one having a physical interest.

We do not think so. On the contrary, notwithstanding its undeniable
importance, we believe that their almost exclusive use in physics,
reinforcing overly the space-time cut into space plus time,
exaggerates the physical interest of the evolution vision (i.e. of
the leading role of  time dependence of spatial configurations in
the description of space-time changes of physical systems).

Other cuts of the space-time may present, and presents, their
intrinsic interest. It is the case, for example, of the Solar system
synchronization, which foliates the space-time by time-like
instants, as we have shown in subsection \ref{subsec5SolarTime}. And
more importantly, also the case of the positioning systems, cutting
any (history of an) extended object by four (histories of)
electromagnetic pulses.

The very concept of synchronization, foliating space-time by
instants not necessarily related to simultaneity, is revealed to be
a gentle but powerful instrument which  allows us to get in training
to `see' space-time under different, unconventional, viewpoints. In
fact, as we have shown in subsection  \ref{subsec5LinSyncGr}, the
simple linear synchronization group is able to already generate
coordinate systems of {\em any} Newtonian class.

Once became used to handle arbitrary synchronizations, one can try
to learn to describe nature without using {\em any} synchronization
at all. This is possible by means of the positioning systems.
Although, of course,  they can be {\em related} to standard
coordinate systems, they do not {\em contain} their causal specific
features, as reveals the fact that they necessarily belong to {\em
any} of the other three classes, the standard one being excluded, as
we have seen in subsection \ref{subsec5NewEmission}.

But the role of Newtonian spacetime in modern gravitational physics
is principally that of facilitate the comprehension of the analog
features of the relativistic space-time. Consequently, in section
\ref{sec:6}, we have also considered the relativistic analogues of
the above mentioned  systems.  1) For simplicity we have applied the
local Solar time synchronization to the relativistic  rigidly
rotating disk, and we have seen in subsection \ref{rotation} that,
in the interior of the light cylinder, the resulting causal class
has same causal signature that the Newtonian one. 2) Concerning the
linear synchronization group, we have shown in subsection
\ref{LinSynchroMinko} that it generates $29$ causal classes, four of
them having the same causal signature that the four Newtonian ones.
3) And finally, in contrast to the Newtonian case, we have seen in
subsection \ref{subsec4RelEmission} that positioning systems in
relativity may be of $103$ causal classes, three of them having the
same causal signature that the corresponding  $3$ Newtonian causal
classes, and only one of them, the $\{{\rm e \, e \, e \, e,}$ $
\,{\rm E \, E \, E \, E \, E \, E} ,$ $ \, l \, l \, l \, l \}$
corresponding to relativistic positioning systems based in light
signals.

The ability to take hold of Newtonian space-time without the use of
the simultaneity foliation may seem rather academic. But such
ability for the relativistic space-time seems urgent. Simply
because, in relativity, relative simultaneity foliations, be them
introduced as an approximate concept or as an exact one, have
neither more nor less physical reality than the celestial crystal
spheres of the Ptolemaic epicyclic theory of planets.

Such foliations are conventional constructions whose realization
really  demand the a priori knowledge of (a good number of) the
physical quantities that usually one wants to know. As such
constructions, they can play a role for the `a posteriori' physical
interpretation of some physical quantities, but are  unusable as
{\em  starting} basis for referring  physical observations of a
unknown environment.

The direct confrontation of the physicists with their environment in
order to know it gravitationally is a basic problem still unsolved
in relativity. Such a confrontation needs a locating structure that,
in order to not to chase its tail, be able to be constructed {\em
before} the  measure of the gravitational properties be done. As has
been analyzed elsewhere (see, for example, \cite{2D-A,2D-B,4D}) this
locating structure is constituted by the relativistic positioning
systems broadcasting light signals in vacuum.

\ack This work has been supported by the Spanish Ministerio de
Educaci\'on y Ciencia, MEC-FEDER project FIS2006-06062.

\end{document}